\documentclass[twoside]{ima}
\usepackage{amssymb}

\usepackage{color}\definecolor{DkRed}{cmyk}{0,.5,.5,.4}
\newcommand\Nmarginpar[1]{\-\marginpar{\raggedright\tiny{\color{DkRed} #1}}}
\def\Ncut#1{{\color{blue}\footnotesize [[#1]]}} 
\def\Nfix#1#2{{\bfseries\slshape #1}}
\def\Nmemo#1{\Nmarginpar{PN:#1}}

\def\altfig{n}  
\newcommand{\cut}[1]{}
\def\Rcut#1{} 
\def\Rmemo#1{\Nmarginpar{RP:#1}}

\def\Jcut#1{{\color{blue}\footnotesize [[#1]]}} 
\def\Jfix#1#2{{\bfseries\slshape #1}}

\def\Jcut{}\def\Ncut{}\def\Nfix#1#2{#1}\def\Jfix#1#2{#1}\def\Nmemo#1{}\def\Rmemo#1{}

\def\NRbead{R_{\rm bead}}
\def\Jrmsfour{\text{RMS}_{4\,\mathrm s}}
\def\Jrmst{\text{RMS}_{t\,}}

\usepackage{amsmath}

\usepackage{graphicx}

\def\figtest{y}
\def\eref#1{Eq.~\ref{#1}}

\def\fref#1{Fig.~\ref{#1}}\def\Fref#1{Fig.~\ref{#1}}
\def\rref#1{Ref.~\cite{#1}}

\newcommand{\ex}[1]{{\mathrm e}^{#1}}                 

\def\msunit{\ensuremath{\mathrm{msec}}}
\def\sunit{\ensuremath{\mathrm{s}}}

\def\nmunit{\ensuremath{\mathrm{nm}}}

\newcommand{\av}[1]{\langle #1\rangle}

\title{CALIBRATION OF TETHERED PARTICLE MOTION EXPERIMENTS}

\author{LIN HAN\thanks{Department of Applied Physics, California Institute of Technology, Pasadena CA 91125. Partially supported by
the Keck Foundation, National
Science Foundation grants CMS-0301657 and CMS-0404031, and the
National Institutes of Health Director's Pioneer Award grant  DP1
OD000217.}
\and
BERTRAND H. LUI$^{*}$\thanks{Current address: Department of Bioengineering, Stanford University, Stanford, CA.}
\and
SETH BLUMBERG$^{*}$\thanks{Current address: University of Michigan Medical Scientist Training Program, Ann Arbor, MI 48109.}
\and
JOHN F. BEAUSANG\thanks{Department of Physics and Astronomy, University of  Pennsylvania, Philadelphia PA 19104. Partially supported by
NSF grants DGE-0221664, DMR04-25780, and DMR-0404674.}
\and
PHILIP C. NELSON$^{\S}$
\and
ROB PHILLIPS$^{*}$\thanks{Corresponding author: {\tt phillips@pboc.caltech.edu}.}
}

\begin{document}

\maketitle


\begin{abstract}
The Tethered Particle Motion (TPM) method has been used to observe and characterize a variety of protein-DNA interactions including DNA looping and transcription. TPM experiments exploit the Brownian
motion of a DNA-tethered bead to probe biologically relevant conformational changes of the tether. In these experiments, a change in the extent of the bead's random motion is used as a reporter of the
underlying macromolecular dynamics and is often deemed sufficient for TPM analysis. However, a complete understanding of how the motion depends on the physical properties of the tethered particle
complex would permit more quantitative and accurate evaluation of TPM data. For instance, such understanding can help extract details about a looped complex geometry (or multiple coexisting
geometries) from TPM data. To better characterize the measurement capabilities of TPM experiments involving DNA tethers, we have carried out a detailed calibration of TPM magnitude\Ncut as a function of DNA length and particle size\Jcut. We also explore how experimental parameters such as acquisition time and exposure time
affect the apparent motion of the tethered particle. We vary the DNA length from 200\thinspace{}bp to 2.6\thinspace{}kbp and consider particle diameters of 200, 490 and 970\thinspace{}nm. We also
present a systematic comparison between measured particle excursions and theoretical expectations, which helps clarify both the experiments and models of DNA conformation. \\
\medskip
\end{abstract}

{\AMSMOS Primary 92C05
; secondary 92C40
, 92C37.
\endAMSMOS}

\newpage\section{Introduction}
\begin{figure}
   \begin{center}
\ifx\altfig\figtest
       \includegraphics*[width=4.4truein]{calibrationfigPN/lac.eps}\\
\else
       \includegraphics*[width=4.4truein]{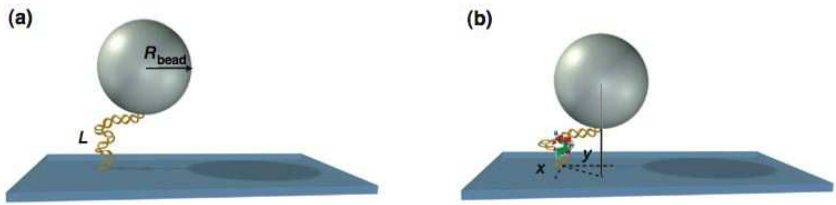}\\
\fi
      \caption{\small\label{fig:TPM} Idea of the tethered particle motion method. Cartoons showing  the tethered bead in the (a) absence and (b) presence of a DNA-binding protein, which changes the
      effective tether length by looping and/or bending the DNA. For example, Lac repressor protein (LacR) has two binding sites, which recognize and bind to two specific sequences (``operators'') on 
DNA.
\Ncut}
   \end{center}
\end{figure}

\let\citep=\cite
\Nmemo{Before submission: Check no figure wider than 4.48 inches. Spell check. Remove doublespacing etc. Fix any overfull lines.
}Single molecule studies are enriching our understanding of biological processes by providing a unique window on the micro\-trajectories of individual molecules rather than their
ensemble-averaged behavior. Many of these studies are devoted to exploring the intricacies of protein--DNA interactions that are central to gene regulation, DNA replication and DNA 
repair\Rmemo{cite van Oijen, Wuite\ldots}.
The resolution of nanometer-scale distances involved in such interactions poses a significant challenge. The emergence of the tethered particle motion (TPM) method offers a practical and relatively
simple solution. In this method, a biopolymer is tethered between a stationary substrate and a micrometer-scale sphere (a ``bead''), which is large enough to be imaged with conventional optical microscopy
(\fref{fig:TPM}). The constrained Brownian motion of the bead serves as a reporter of the underlying macromolecular dynamics, either by observing its blurred image in a long exposure \citep{Finzi1995}, or by
tracking its actual trajectory in time (e.g.\ as done in \citep{Pouget2004} and the present work). Changes in the extent of the motion (which we will call ``excursion'') reflect conformational transformations of the tethered molecule. Such changes
may be caused by processive walking of RNA polymerase~\citep{Schafer1991,Yin1994}, DNA looping~\citep{Finzi1995, Broek2006, Zurla2006,Zurla2007,Wong2007,Vanzi2006,Beausang2007}, DNA hybridization
\citep{Zocchi2003}, DNA bending \citep{Toli2006}, Holliday junction formation \citep{Pouget2004} or RNA translation \citep{Vanzi2003}.\Rmemo{Include Salome, van Oijen, Wuite on restriction enzymes. 
Add action o the enzymes of DNA replication and recombination (van Oijen, Gelles on integrase).}

Although TPM is simple in principle, there are a variety of technical challenges that must be addressed for successful implementation. For example, sample preparation can be compromised by
multiply-tethered beads, non-specific adsorption, transient sticking events and dissociation of the tether joints \citep{Pouget2004, Vanzi2006,Broek2006, Blumberg2005, Nelson2006}. In addition, image
analysis of TPM data is complicated by instrumental drift and the stochastic nature of the tethered particle's motion. Several time scales must  be considered, including the total observation time,
exposure time, and the intrinsic diffusive time scale of the tethered particle. We will show that quantification of the spatial and temporal resolution of TPM measurements requires an understanding of how
particle motion depends on tether length, particle size and other controllable parameters.
\Ncut%
We focus exclusively on TPM behavior in the absence of externally applied force (as might be applied via magnetic or optical tweezers). 

The aims of this article are to: (1) review how data acquisition and data analysis affect TPM measurements; (2)~explain a practical scheme of data selection and quantify the 
fractions of typical data that are rejected by each of our criteria; (3)~calibrate particle motion, tether length and observation time so
that subsequent TPM experiments can be quantitatively interpreted; and (4) discuss the physical processes that govern TPM. Calibration of the particle motion allows precise predictions of how a
particular conformational change of the tether, such as Lac repressor induced looping of DNA, affects TPM.

Some of our experimental results were outlined in \citep{nels07a}. Theoretical work leading up to the present results on TPM motion appeared in \Ncut\citep{sega06a,Nelson2006,Towles2008}. For 
example, Segall et al.\ predicted effects of changing the size of the bead and tether length, which we document experimentally in the present work. Our 
results are preparatory to experimental \citep{han?cLength} and theoretical \citep{Towles2008} work on DNA looping in the \textit{lac} operon system.
 


\section{Results and Discussion}
Using differential interference contrast (DIC) microscopy, the projected position of several beads in a field of view are recorded using a CCD camera. \Jfix{Sub-pixel resolution}{} position traces for
each bead in the image is determined using a cross-correlation method \citep{Gelles1988}. Standard microscopy systems such as ours are limited to two spatial dimensions; tracking of three
dimensions has been accomplished using evanescent fields or diffraction rings \citep{Blumberg2005}, but this involves additional calibration and technical challenges. Two-dimensional tracking is 
sufficient for the applications we have in mind, such as DNA-looping studies. The tracked position of the bead is
subject to slow drift, due to vibrations of the experimental apparatus, which we removed using a first order Butterworth filter at 0.05\thinspace{}Hz cutoff frequency \citep{Vanzi2006}. To
quantify bead excursion, we then used the square root of the sum of the variances of the drift corrected particle position $(x, y)$ along two orthogonal image-plane axes:
\begin{equation}
\text{RMS}_{t} = \sqrt{\av{(x-\bar{x})^2 + (y-\bar{y})^2}_{t}}
.\label{eq:rms}
\end{equation}
Here $t$ is the time interval over which the RMS motion is measured (typically 4\thinspace{}s); $\bar{x}$ and $\bar{y}$ represent the average of $x$ and $y$ over time $t$. \eref{eq:rms} is evaluated 
as a sliding filter at each point along the trajectory, and permits us to capture the tether dynamics using a single scalar quantity, as illustrated in \fref{fig:loopsensitivity} below. The 
finite-sample means $\bar x, \bar y$ are subtracted as an additional method of eliminating instrumental drift not removed by the Butterworth filter; in practice, this subtraction has little effect. 
When simulating the motion numerically, we will compute the same quantity as \eref{eq:rms}, in order to make an appropriate comparison.

\subsection{Data selection criteria}
Although single-particle tracking data can reveal detailed features of the dynamics of protein-DNA interactions, care must be taken to minimize experimental artifacts such as non-specific binding of the bead
and DNA to each other and the surface, as well as multiple DNA attachments on the same bead. To get acceptable calibration data, we implemented several selection criteria called ``minimum motion,''
``motion symmetry,'' and ``uniformity.'' \Ncut

\begin{figure}[htp]
  \begin{center}
\ifx\altfig\figtest
\includegraphics*[width=4in]{calibrationfigPN/selectionrules.eps}\\
\vspace{.1in}
    \includegraphics*[width=3in]{calibrationfigPN/selectioneff.eps}
\else\includegraphics*[width=4in]{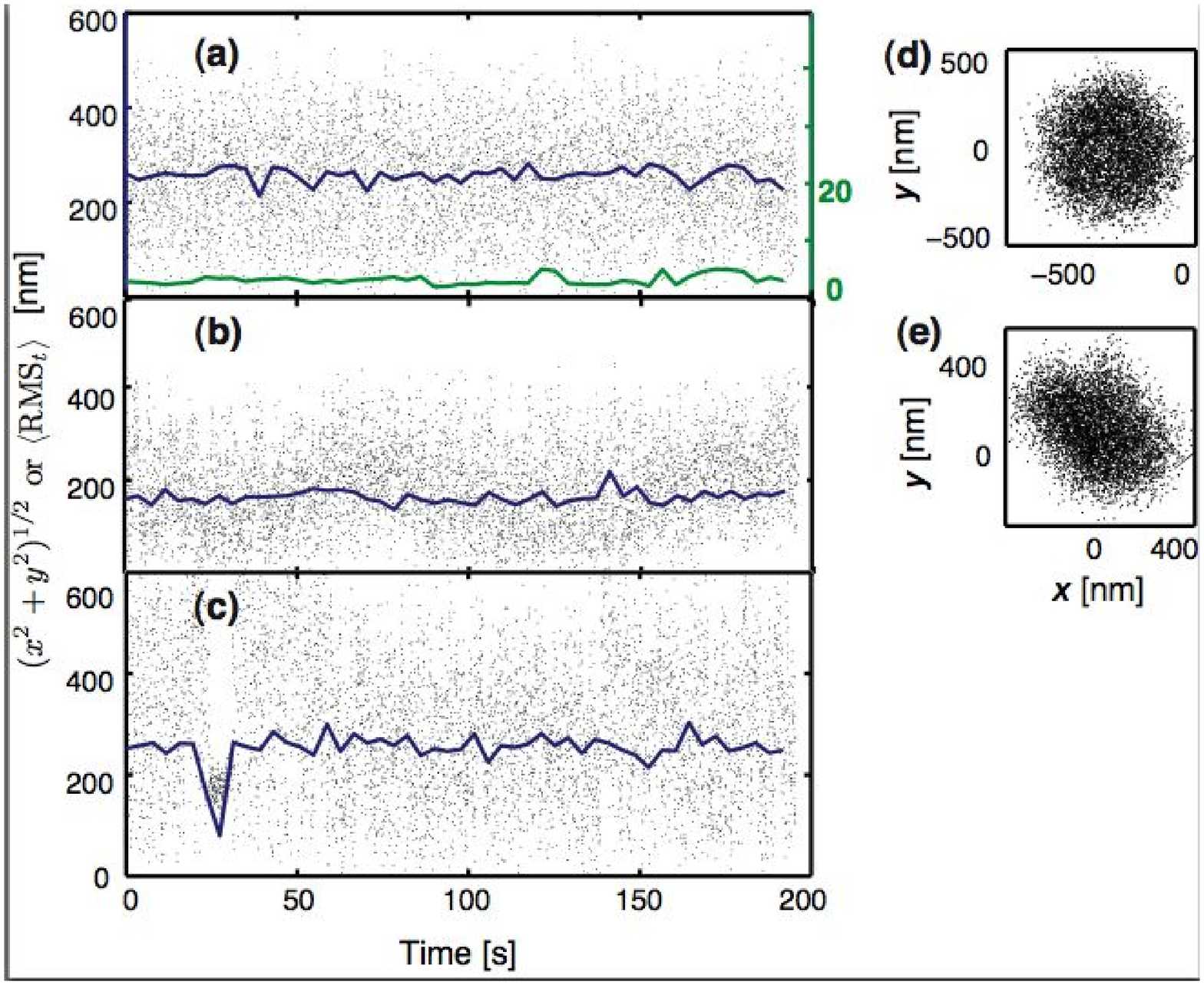}\\
\vspace{.1in}
    \includegraphics*[width=3in]{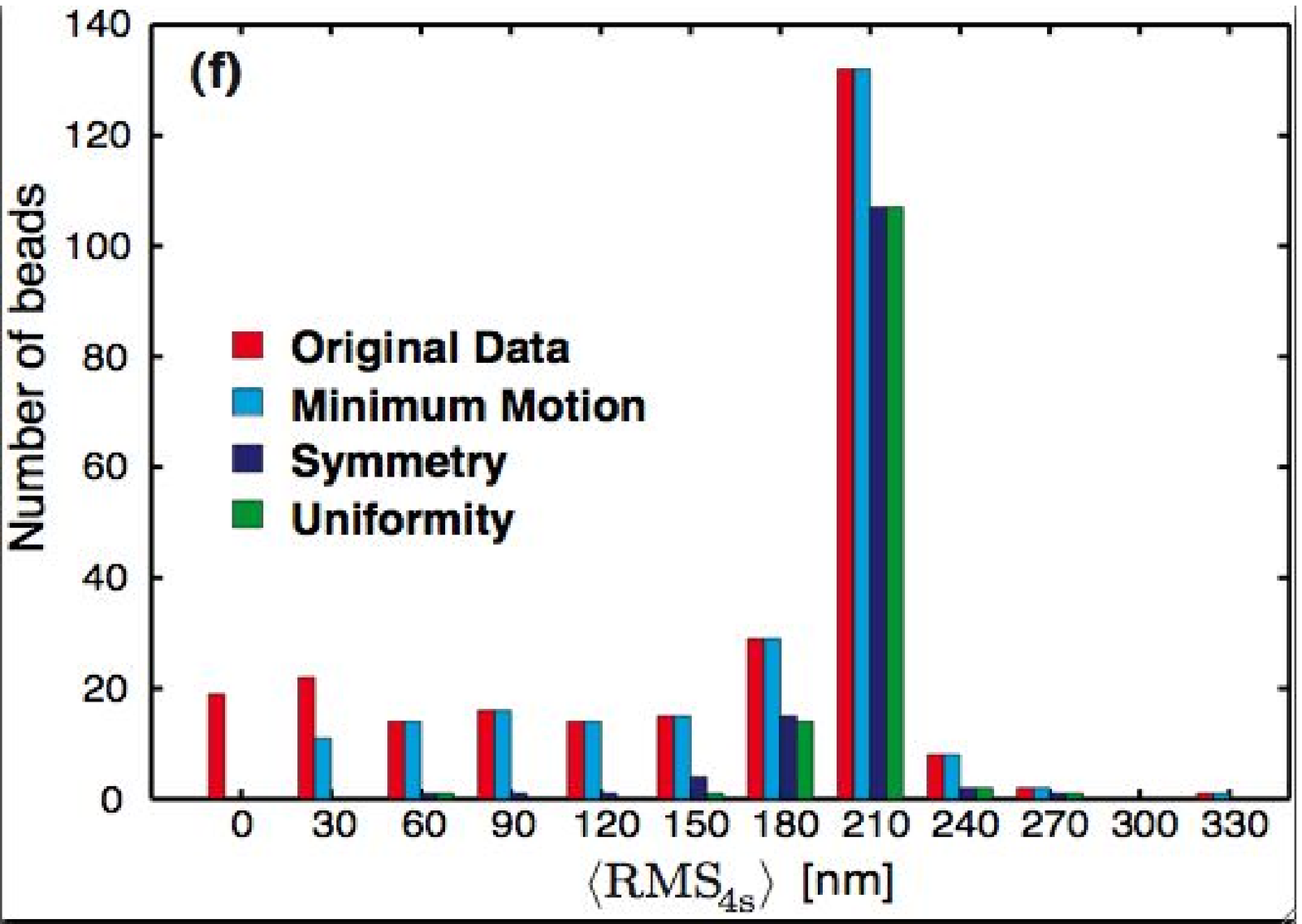}
\fi    \caption{\small \label{fig:selection}
Selection of qualified tethers. In (a--e), the dots \Nfix{show instantaneous positions after drift subtraction}{
 correspond to raw data}; the lines show $\Jrmsfour$. (a) Trajectory associated with an accepted data set (blue) and stuck bead (green). (b)
Trajectory for a bead that passed ``minimum motion'' but failed the ``motion symmetry'' test (see (e)). (c) Trajectory associated with nonuniform motion caused by transient, nonspecific binding, seen as a downward spike between 0 and 50 seconds. (d)
$xy$ scatter plot of the trajectory in (a) shows it to be symmetric. (e) Scatter plot of the motion in (b) shows it to be asymmetric. The DNA used in (a--e) are 1206\thinspace{}bp long and the bead size is
490\thinspace{}nm in diameter. (f) Distribution of bead  excursions and the number of beads that pass successive application of the
selection criteria (see text) \cite{Blumberg2005}.
Red:     original data.
Cyan:    after application of minimal motion filter.
Blue:    after application of symmetry filter.
Green:    after application of uniformity filter.   The DNA used in (f) are 901\thinspace bp long and bead size is 490\thinspace{}nm diameter. }
  \end{center}
\end{figure}
``Minimum motion'' discards beads that cannot be differentiated from beads stuck to the glass substrate (or otherwise compromised in their mobility).  \Jcut$\Jrmsfour$\,  from a control experiment with beads but no DNA is shown in \fref{fig:selection} (green line), and is 
substantially smaller than that for a tethered bead (blue line). Data sets exhibiting average excursions, $\Jrmsfour$, lower than 30\thinspace{}nm cannot be differentiated from stuck beads and are therefore rejected.

``Motion symmetry''  requires that  a tethered particle should exhibit symmetric in-plane motion about its anchor point, and is calculated from the covariance matrix \citep{Blumberg2005,Nelson2006}:
\begin{equation}
C= \begin{pmatrix} \sigma_{x_{1}x_{1}} & \sigma_{x_{1}x_{2}} \\
\sigma_{x_{2}x_{1}} &
\sigma_{x_{2}x_{2}}  \\
\end{pmatrix},\label{eq:cov}
\end{equation}
where
\begin{equation}
\sigma_{x_{i}x_{j}} = \frac{1}{N}\sum_{k=1}^{N} x_{i}^kx_{j}^k - \bar x_i\bar x_j 
\end{equation}
are the second moments of the bead's position. Here $N$ is the number of video frames and $x_{1}^{k}$, $x_{2}^{k}$ are the in-plane coordinates (i.e.\ the position $x,y$) of the microsphere for frame
$k$ as obtained from the \Jfix{drift-corrected}{} data. The eigenvalues ($\lambda_1, \lambda_2$) of the covariance matrix indicate the squares of the major and minor axes corresponding to the in-plane
displacement of the bead and are equal for a perfectly symmetric motion. We took $s=\sqrt{{\lambda_{\rm max}}/{\lambda_{\rm min}}} \leq 1.1$ as our  acceptable threshold. \Ncut\Fref{fig:selection}(d,e) displays scatter plots for the in-plane motion of two beads to illustrate the distinction between symmetric and asymmetric
tethers. The first plot passes the symmetry test and would serve as a qualified tether; the second would be rejected. Asymmetric bead trajectories may be caused by multiple DNA
tethers~\citep{Pouget2004}.

``Uniformity''  qualifies tethers on the basis of the consistency of their motion over time and eliminates beads showing non-specific binding events, such as  binding of DNA to the bead or glass surface for short
periods.  To detect these events automatically, we refine a procedure used in \rref{Blumberg2005}. We
first divide the entire time series into 10 subsets labeled by $i=1,\ldots10$. In subset $i$, we calculate $\Jrmsfour$ over each 4\thinspace s 
window and then average these, defining $A_{i}\equiv\langle\text{RMS}_\mathrm{4\,s}\rangle_i$. Then we define $u$ as the standard deviation of $\{A_1,\ldots A_{10}\}$, normalized by the overall 
average $\Jrmsfour$. Only data sets with relative standard deviation $u<0.2$ are accepted.
For example, the bead shown in \Fref{fig:selection}(c) meets the motion and symmetry criterions; however, it displays a non-specific binding event at $~30\sunit$. 
In short, our third criterion removes tethers with temporal inconsistency in their Brownian motion.

The first two selection criteria discard tethers that are permanently defective, whereas the third eliminates time series with undesirable transient events. Note that if the purpose of the
experiment is to identify interesting molecular binding events, such as those leading to DNA looping or bending, then the last criterion cannot be applied, because these transient events can appear
similar to the sticking events rejected by the uniformity criterion.  In the
present work we aimed at characterizing uniform DNA tethers, so we enforced all three criteria. Prior to
applying the selection criteria, \fref{fig:selection}(f) displays a broad distribution in the measured $\Jrmsfour$ (red). Afterwards, $\sim50\%$ of the data are qualified and exhibit well-defined
Brownian motion (green bars). This figure shows that the primary cause of bead rejection is asymmetric in-plane motion. Experimentally, beads with multiple tethers can be minimized by reducing the
concentration of DNA.

\subsection {Acquisition Time} The drift-corrected $(x,y)$ trajectories are  noisy due to the stochastic Brownian motion of the particle, and are thus filtered using \eref{eq:rms} over a
particular time window $t$ (usually four seconds). Although analysis methods exist that make no use of this windowing step \citep{Beausang2007,beau2007}, nevertheless many experiments do use it,
and so we investigated its effect on reported bead excursion. Too short a window will increase the noise, 
leading to broad peaks in the distribution of $\Jrmst$ that make signals from differently sized tethers too difficult to
distinguish. Moreover, for short $t$ the bead will not adequately explore its 
full range of accessible configurations, leading to an underestimate of $\Jrmst$, as we document below. At the other extreme, however, too long a window will result in a loss of temporal resolution.
\Jcut

\begin{figure}[ht]
  \begin{center}
\ifx\altfig\figtest\includegraphics*[width=4.4truein]{calibrationfigPN/winsize-motionallclean1.eps}
\else\includegraphics*[width=4.4truein]{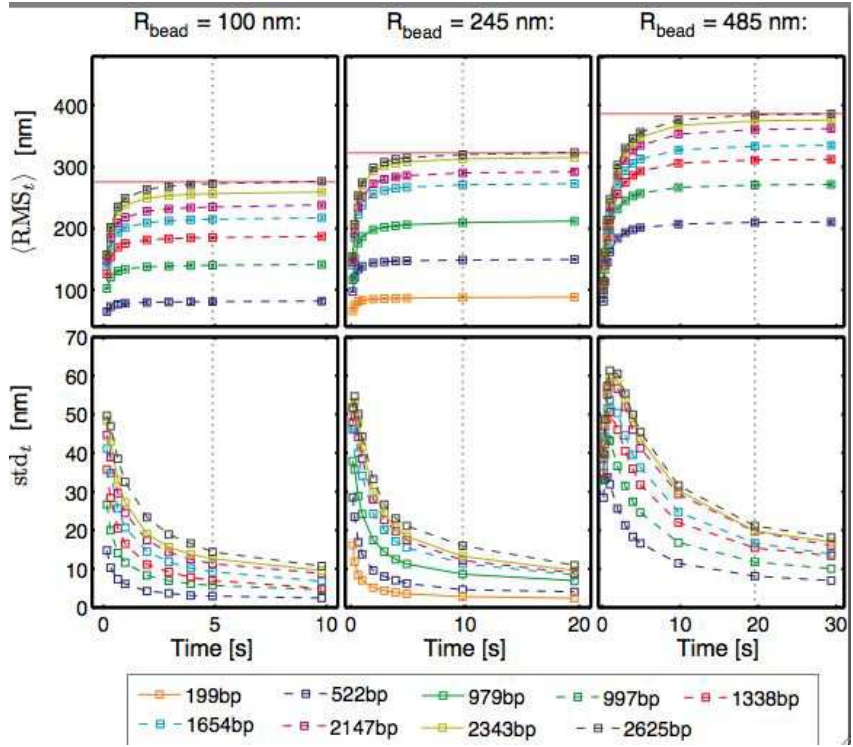}
\fi \caption{\small \label{fig:winsize} \textit{Top:} Average RMS excursion and \textit{bottom:} standard deviation of $\Jrmst$ as functions of window time $t$ in \eref{eq:rms} for different
bead sizes (columns) and lengths $L$ of the DNA tether (colored lines). \Ncut\Jcut\Ncut\Jcut{}As discussed in the text, 
black dotted lines indicate ``large enough'' choices of $t$\Ncut.}
  \end{center}
\end{figure}

To determine the optimum TPM window size, we recorded data for 200$\,\sunit$, for several bead sizes and a wide range of tether lengths,  then found the mean ($\av{\Jrmst}$) and standard deviation ($\mathrm{std}_t = \sqrt{\langle
\Jrmst^2 \rangle- \langle \Jrmst \rangle ^2}$) of the RMS-filtered trajectory for various values of window size $t$ (see \fref{fig:winsize}). Here $\langle\ldots \rangle$ denotes two averages: (1)~over the 
$(200\,\sunit/t) $ windows that make up each bead's time series, and (2) over nominally identical tethered particles with the same bead size and tether length. The DNA lengths varied from 
199\thinspace{}bp to 2625\thinspace{}bp, and we tested beads with three different diameters: 200\thinspace{}nm, 490\thinspace{}nm and 970\thinspace{}nm.
\Jcut

\Nfix{\fref{fig:winsize} shows the trends as we vary $t$, $\NRbead$, and tether length $L$. We first notice that for fixed $\NRbead$ and $L$, each curve levels off as $t\to\infty$, giving an asymptote that 
is the true RMS excursion. (For short times, the bead has not had a chance to explore its full range of motion in any given window, and so each $\Jrmst\to0$, and hence so does $\langle\Jrmst\rangle$.) To make the 
tradeoff discussed earlier, we now ask: How long must we choose the window time $t$ in order to get a reliable estimate of the true excursion?}{}

\Nfix{Naively we might suppose that each video frame gives an independent draw from a distribution of bead positions whose RMS value we seek. In that case, we would expect that as soon as $t/(30\,\msunit)$ 
becomes large, we would have a good estimate of the true RMS excursion. But the top row of \fref{fig:winsize} shows that, on the contrary, the minimum required observation time increases both with increasing 
bead radius (moving between the three panels) and with increasing tether length (moving between the curves on a given panel). Physically, the point is that successive video frames are \textit{not} 
independent draws from the distribution of particle positions, because the particle's motion  is diffusive. The diffusion time $\tau_{\rm diff}$ of a particle in a trap increases with increasing trap 
radius and with increasing viscous drag constant for the particle, giving rise to the trends observed in the figure. (For a theoretical discussion see the Supplement to \citep{Towles2008}.)}{}

\Nfix{Similarly, the second row of graphs in \fref{fig:winsize} shows that the scatter between successive determinations of $\Jrmst$ decreases with increasing $t$. This ``sharpening'' effect also explains 
how RMS filtering takes rather diffuse raw data (e.g.\ \fref{fig:selection}a--c) and transforms it into a fairly well-defined ``state'' (e.g.\ the individual states visible in filtered traces such as 
\fref{fig:loopsensitivity}).
In both rows of \fref{fig:winsize}, we have drawn dotted lines to illustrate a value of $t$ that is ``safe'' (long enough) for tether lengths up to 2600\thinspace{}bp.}{}

\subsection{Calibration of motion}
In order for TPM experiments to detect discrete conformational changes of biopolymers such as in DNA looping, it is necessary to quantify how tether length affects particle motion. Precise calibration
data also indicates the minimum detectable change in tether length. Sensitive measurements may also allow detection of more subtle changes, such as kinking of the DNA upon protein binding or multiple loop
topologies.
\Jcut
\Jcut  

\begin{figure}[ht]
  \begin{center}
\ifx\altfig\figtest\includegraphics*[width=4truein]{calibrationfigPN/cali-curve-eq4sclean1.eps} \else \includegraphics*[width=4truein]{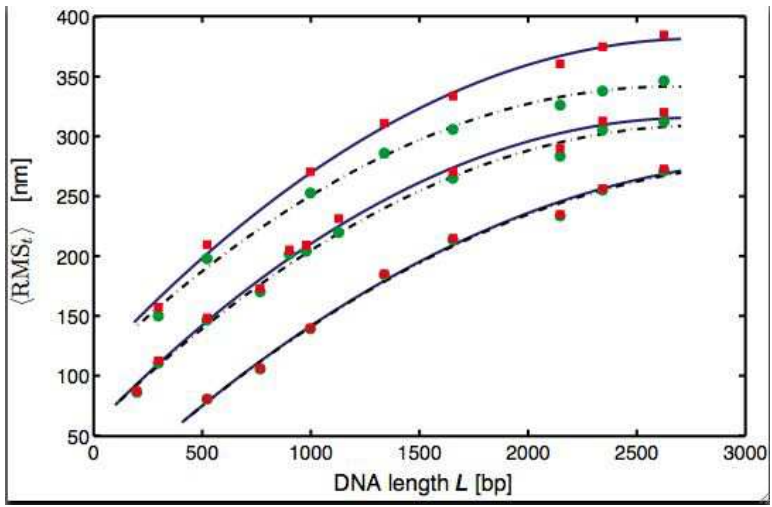} \fi 
\caption{\label{fig:calibration} \small
RMS excursion of bead as a function of the tether length for different sized microspheres, \Nfix{for random-sequence DNAs of various lengths}{}. Each red square is the average of
equilibrium amplitude of RMS motion over 20 to 200 qualified beads, which is calculated by using \eref{eq:rms} with $t = 5\,$s for $R = 100\,$nm (bottom data set), $t = 10\, $s for $R = 245\, $nm
(middle data set) and $t = 20\,$s for $R = 485\, $nm (top data set). Using $t=4\,\sunit$ for the same data systematically underestimates the motion of larger beads (green circles). \Ncut{}The curves are empirical polynomial fits to the datasets\Ncut (see Table~\ref{tab:fit}). }
  \end{center}
\end{figure}
\Ncut

\Nfix{To find the empirical calibration curve, we created many DNA tethers of varying lengths, and attached beads of three different sizes. For each bead size, we estimated the RMS excursion by its
finite-sample estimate $\Jrmst$, taking $t$ to be the lowest ``safe'' value as estimated in the previous subsection: $t= 5\,\sunit$, 10\thinspace{}s and 20\thinspace{}s for beads with diameters of
200\thinspace{}nm, 490\thinspace{}nm and 970\thinspace{}nm respectively, with results shown in \fref{fig:calibration}. (For comparison, we also show corresponding results with $t$ fixed to 
4\thinspace{}s, which deviate significantly from the longer observations for the larger beads.) We summarized all these data with polynomial fits shown in the figure and given explicitly in 
Table~\ref{tab:fit}. }{}

\begin{table*}
\caption{\small Parameters of quadratic function $ax^2+bx+c $ obtained for fitting both the equilibrium motion data (red squares in \fref{fig:calibration}) and 4\thinspace{}s interval data (green
circles in \fref{fig:calibration}).}
\label{tab:fit}
\begin{center}
\begin{tabular}{|ccccc|}
 \hline  Time [s] &Diameter         & $a\times10^{-5}$ &$ b$ &$c$  \\  \hline
    5&200\thinspace nm &  -2.58 $\pm$ 0.68     &0.17 $\pm$ 0.02   & -4.5$\pm$14.8 \\
   10&  490\thinspace nm & -3.37 $\pm$ 0.47   &  0.19 $\pm$ 0.01  &   57.3$\pm$ 7.2 \\
   20&  970\thinspace nm & --3.49 $\pm$0.46    & 0.20 $\pm$ 0.01 &  109.5 $\pm$ 8.7\\   \hline
      4&  200\thinspace nm &  -2.60  $\pm$0.69     &0.17 $\pm$0.02   & -4.75$\pm$14.7\\
    4& 490\thinspace nm & -3.17 $\pm$ 0.41  &  0.18$\pm$ 0.01  &   58.05$\pm$ 6.6 \\
   4&  970\thinspace nm & --3.31 $\pm$0.48   & 0.18 $\pm$ 0.01 &  107.7 $\pm$ 8.7\\   \hline

\end{tabular}
\end{center}
\end{table*}

\subsection{Theoretical predictions}
\begin{figure}[ht]
  \begin{center}
\ifx\altfig\figtest\includegraphics*[width=3.54truein]{calibrationfigPN/KBT_calib.eps}
\else\includegraphics*[width=3.54truein]{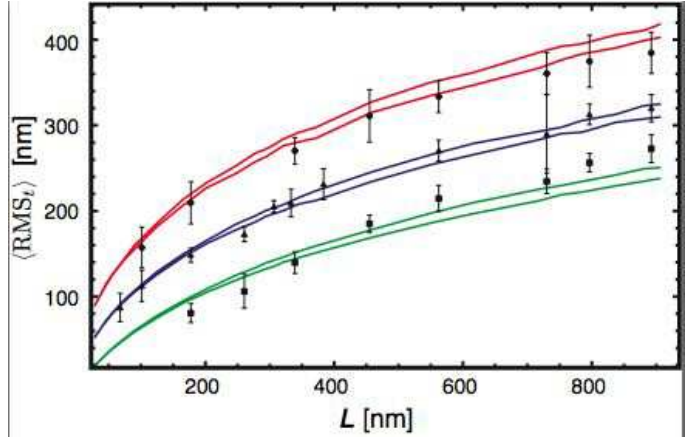}
\fi \caption{\small \label{calibrplot}  Theoretical prediction of equilibrium bead excursion, following a method introduced in \cite{sega06a,Nelson2006}.
\textit{Dots:} Experimental values (same as red squares in \fref{fig:calibration}).
Each dot represents $20$--$200$ different observed
beads, with the given tether length. Each such bead was observed for about $200\,\sunit$, yielding $(200\,\sunit)/t$ measurements of the RMS motion, which were averaged; here $t=20,\ 10$, and $5\,$s 
as in \fref{fig:calibration}. Each data point shown is
the average of these averages; error bars represent the variation (standard deviation) among the beads. 
\textit{Curves:} Theoretically predicted RMS motion, corrected for the blurring effect of finite shutter time. For each of the three bead sizes studied, two curves are shown. From top to
bottom, each pair of curves assumes persistence length values $47\cut{}$ and $39\,\nmunit$, respectively, a range appropriate for the solution conditions we used \citep{wang97a}. 
There are \textit{no fit parameters;} the theoretical model uses values for bead
diameter given by the manufacturer's specification. The bumpiness in the curves reflects the statistical character of the Monte Carlo algorithm that generated them.
}
  \end{center}
\end{figure}
We also compared the experimental data in \fref{fig:calibration} to a mathematical simulation of the bead-tether-wall system (\fref{calibrplot}). The excursion of the bead away from its attachment
point on the microscope slide is affected by the length and stiffness of the DNA tether, the size of the bead, and the various interactions between the bead/wall, bead/tether, and wall/tether. To
account for all these effects, we modified the Gaussian sampling Monte Carlo technique previously used in \cite{sega06a,Nelson2006,czap06a, nels07a} (see \citep{Towles2008} for details).  

Suppose first that a semiflexible polymer chain is anchored at one point in space, but is otherwise unconstrained. At the anchored point we suppose we are given a probability distribution of 
different possible initial orientations for the first chain segment. The distribution of positions and orientations of the other end is then a convolution of this initial distribution with a kernel 
representing a particular diffusion process (random walk) on the group manifold of the three-dimensional Euclidean group. 

We can numerically compute moments of this final distribution, or its various marginal distributions, by a Monte Carlo procedure. Idealizing the polymer as a chain of finite elements, each is related 
to its predecessor by a shift along the latter's 3-axis, a twist about the same axis, and some random bend and twist. Rather than represent the random part using Euler angles, a more invariant 
formulation is to draw a $3\times3$ generator matrix from a Gaussian distribution on the Lie algebra so(3), then exponentiate it. The Gaussian distribution is determined by a covariance matrix, which 
represents the bend and twist elasticity of the DNA, together with bend-twist couplings. We estimated it up to an overall rescaling factor from structural data on DNA, then chose the overall factor 
to yield a desired value of the persistence length of DNA.

Turning from the idealized problem above to TPM, we see that we must implement steric constraints: One end of the DNA tether is attached to a wall, which the DNA may not penetrate. Moreover, the 
other end is attached to the sphere, which itself must not penetrate the wall. Nevertheless, each segment of the intervening DNA is otherwise free to bend, independently of its neighbors. Thus the 
same Monte Carlo generation just described continues to be valid, except that some sterically forbidden chains must be discarded. Thus
our 
computer code generated many simulated DNA chains and bead orientations in a Boltzmann distribution, applied the steric
constraints \cite{sega06a}, and tabulated the resulting values of the distance from the projected bead center to the attachment point. The necessary calculations were coded in \textsl{Mathematica} 
and ran conveniently on a laptop computer.

We chose to
compare to experimental data with ``safe'' values of the window time $t$, so we simply had the code evaluate the RMS value of this distance. (For a procedure valid for any $t$, see the Supplement to 
\citep{Towles2008}.)
We also applied a correction to this theoretical result, to account for the bead's
motion during the rather long shutter time (see the following subsection). \fref{calibrplot} shows that an \textit{a priori} calculation of the expected motion matches the data fairly well, with a
value of persistence length consistent with others' experiments; there were no other fitting parameters. \Ncut

\subsection{Blurring Effect}
\begin{figure}[ht]
  \begin{center}
\ifx\altfig\figtest\includegraphics*[width=3.1truein]{calibrationfigPN/KBT_blur.eps}
\else\includegraphics*[width=3.1truein]{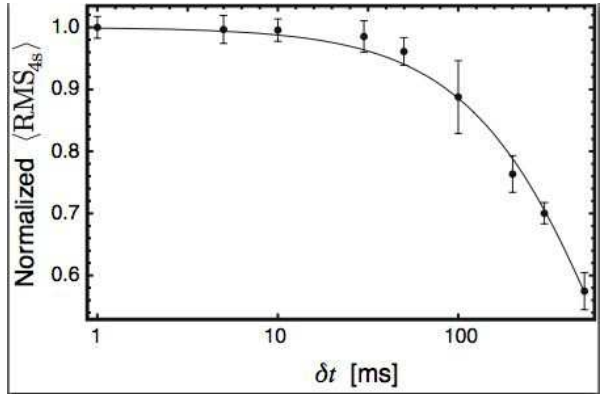}
\fi    \caption{\small \label{fig:exposure}{RMS bead excursion as a function of camera shutter time in milliseconds. \textit{Dots:} Experimental data.  Each dot represents about $20$ 
different observed beads, with a tether of length $L=901\,$bp and a 490$\,\nmunit$ diameter bead. Error bars were drawn using the same method as in \fref{calibrplot}.
Each point has been normalized to the data at 1\,ms to give a dimensionless quantity on the vertical axis. \textit{Curve:} Expected correction due to finite shutter speed, calculated by the method in the text 
(\eref{e:1a}), with shutter time given on the horizontal axis (see also the Supplement to \citep{Towles2008}).
}}
  \end{center}
\end{figure}
In our experiments the camera had a long shutter time ($\delta t=31\,\msunit$). During each exposure, the bead moved, creating a blurred image whose center is not quite the same as the instantaneous center.
This blurring effect reduces the apparent bead excursion. Suppose for example that the bead has a momentary excursion to a large value of $x$. Subsequently, its stretched tether will pull it inward,
so that the average position during the video frame has a smaller value of $x$. We quantified this effect using a 901\thinspace{}bp DNA and a 490\thinspace{}nm diameter bead at 1, 5, 10, 30, 50, 100,
200, 300 and 500$\, \msunit$ exposures (\fref{fig:exposure}). Longer exposures indeed reduce the apparent RMS motion of the bead. The effect is minimal for exposure times smaller than
30\thinspace{}ms, but decreases sharply above this value.

These effects can be considered from a theoretical perspective (see the Supplement to \citep{Towles2008}). The effect of the tether on the bead may be approximated as a harmonic restoring force. If 
the bead starts at a distance $\rho_0$ from the center, then its average position drifts inward under the influence of this force. Averaging that trajectory over the video frame gives a blurred 
trajectory with center at $S(\rho_0)\rho_0$, where the blur factor is  
\begin{equation}S(\rho_0)=\frac{T_{s}}{\delta t}[1-\ex{-\delta t/T_{s}}]\label{e:1a}\end{equation}
The time constant $T_s$ can in principle be estimated from first principles, but in practice we fit it to data such as those in \fref{fig:exposure}.
For very small $\delta t$ we get $S\to 1$. For large $\delta t$, we have $S\to0$. 

To predict the experimental data we should thus take the theoretical prediction and correct it by a factor of $S$. This correction is trivial to apply (comes out of
the statistical averaging), because $S$ is independent of $\rho_0$. The curves in \fref{fig:exposure} show that a correction of the form of \eref{e:1a} fits the data well; this correction was applied 
when drawing the curves in \fref{calibrplot}.

\section{Applications to DNA Looping}
\begin{figure}[ht]
 \begin{center}
\ifx\altfig\figtest\includegraphics*[width=4truein]{calibrationfigPN/sensitivity_comp.eps}
\else\includegraphics*[width=4truein]{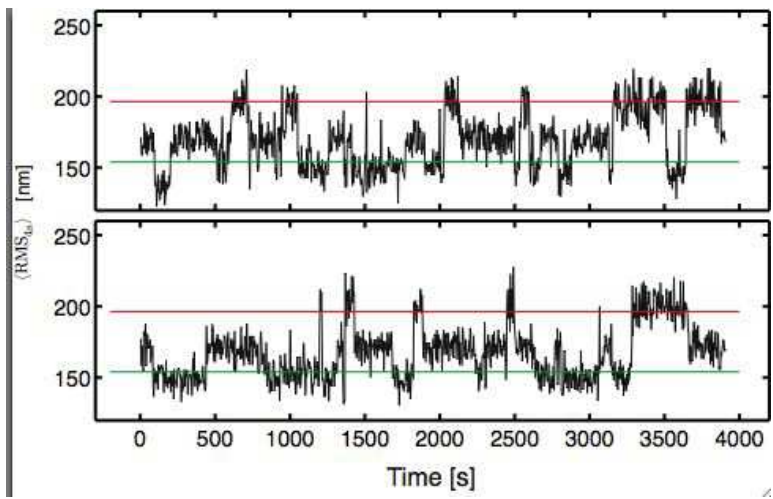}
\fi    \caption{\small \label{fig:loopsensitivity} Two typical $\Jrmsfour$ trajectories  in the presence of Lac
    repressor,  showing events of loop formation and breakdown\Ncut.  \Ncut{}Total length of the DNA tether is $L=901\,$bp\Nfix{; bead diameter is 490\,nm.}{}
\Ncut{}Operator 
center to center distance is 325.5\thinspace{}bp\Ncut. \Nfix{The upper horizontal line is the expected excursion from the calibration curve for the full tether; the 
lower horizontal line is the expected excursion for a tether of length 901-325\,bp.}{}}
 \end{center}
\end{figure}
One of the key applications of the tethered particle method has been its use in
studying DNA looping.  Many transcriptional regulatory motifs involve the binding
of transcription factors that bind at more than one site simultaneously, forming
a loop of the intervening DNA (\fref{fig:TPM}).  The TPM technique has been used to explore these problems.
The calibration analysis performed here can serve as the basis of a more careful
evaluation of DNA looping and bending by DNA-binding proteins, \Ncut%
and a guide to optimize the design of subsequent DNA looping experiments. For example, one may ask, what is the optimal total DNA length and bead size needed to reliably detect a particular type of loop? 
To answer such questions, first note that \fref{fig:winsize} quantifies how smaller beads and shorter tethers both allow us to work with small window size $t$, while still giving the narrow peak widths 
necessary to resolve substructure in the distribution of $\Jrmst$.  \fref{fig:calibration} reinforces this point and also quantifies how smaller beads also optimize the resolution of TPM by 
maintaining a high slope to the calibration curve over a wide range of $L$. There are limits to what can be achieved in this way, of course: 
Small beads are hard to observe, and short DNA tethers tend to collapse (due to surface absorption). Our work helps the experimenter to make appropriate tradeoffs when designing experiments.

\Ncut

Another benefit derived from the calibration curve\Ncut{} is a better understanding of the geometry of the conformational changes we have studied. \Ncut%
\Nfix{For example, \fref{fig:loopsensitivity} clearly shows the existence of a \textit{third} state, not coinciding with either of the horizontal lines naively predicted from the calibration curve 
\cite{Wong2007,norm08a,han?cLength}. More detailed simulations can then shed light on the geometries of the two distinct looped species disclosed by TPM assays \cite{Towles2008,han?cLength}.
}{}

\section{Conclusions}
The tethered particle motion method is one of the simplest tools for performing
single-molecule experiments on DNA-protein complexes.  In contrast to other methods involving fluorescence, TPM never bleaches, allowing very long observations.
The central idea is to use the Brownian motion of  a small particle tethered to a DNA molecule as a reporter of the underlying macromolecular dynamics of the DNA in its complexes with DNA-binding
proteins.   The point of this paper has been to examine the challenges that are inherent
in making useful quantitative measurements using this method.     One of the
main outcomes of that effort has been the development of calibration curves that
illustrate how tethered-particle excursions depend upon both bead size and tether length.

\section{Materials and Methods}

\subsection{Sample preparation}
The first step in any TPM experiment is construction of the relevant DNA tethers with their associated reporter beads.  Polymerase Chain Reaction (PCR) was used to amplify labeled DNA with two modified primers. The primers were either biotin or digoxigenin labeled at the 5' ends (MWG Biotech AG, Ebersberg, Germany).  The labels permit specific linkage of the DNA to a polystyrene microsphere or glass coverslip, respectively.  The PCR templates were taken from lambda phage or modified pUC19 plasmid (sequences available upon request). The PCR products were purified by gel extraction (QIAquick Gel Extraction Kit, QIAGEN).

Streptavidin (Bangs lab) or neutravidin (Molecular Probes) coated microspheres of diameter 200, 490 and 970\thinspace{}nm served as our tethered 
particles.  In contrast to the 490  and 970\thinspace{}nm  microspheres, the 200\thinspace{}nm microspheres were fluorescent. Prior to incubation with DNA, a buffer exchange on the beads was performed by three cycles of centrifugation and resuspension in TPB buffer (20mM Tris-acetate, pH=8.0, 130mM KCl, 4mM MgCl$_{2}$, 0.1mM DTT, 0.1mM EDTA, 20 $\mu$g/ml acetylated BSA (Sigma-Aldrich), 80 $\mu$g/ml heparin(Sigma-Aldrich) and 3 mg/ml casein (Sigma), filtered with 300kD MWCO polysulfone membrane (Millipore)).  This combination of reagents was chosen in an attempt to maximize sample yield and longevity, while minimizing non-specific adsorption of DNA and microspheres onto the coverslip.

The second step is DNA tether assembly.  Tethered particle samples were created inside a 20-30\, $\mu l$ flow cell made out of a glass slide, glass coverslip, double-sided tape and tygon tubing.  The coverslip and slide were cleaned with 4N HCl for 24 hours and then the flow cell was constructed in the same manner as described by van Oijen  {\it et  al.}~\citep{vanOijen2003}.  Next, the flow chamber was incubated with 20 $\mu$g/mg anti-digoxigenin (Sigma) in PBS buffer for 30 minutes, and then rinsed with 400 $\mu$l wash buffer (TPB buffer with no casein)  followed by 400 $\mu$l of TPB buffer.  Microsphere-DNA complexes were created by incubating approximately 100 pM microspheres  with 10 pM labeled DNA in TPB buffer for at least an hour. The DNA concentration was estimated via gel band strength.  The 10:1 ratio of beads to DNA was designed to minimize the occurrence of multiple DNA strands attached to a single microsphere.  The tethering procedure was completed by introducing 50 $\mu$l of the microsphere-DNA complexes into the flow cell for four to ten minutes.  Additional tethering yield could be accomplished by another round of incubation with fresh microsphere-DNA complexes.  Finally, unbound microspheres were removed by flushing the chamber with 1 mL TPB buffer. Once microspheres were introduced into the flow cell, tether integrity was improved by taking care to minimize flow rates within the sample chamber.
%

\subsection{Data Acquisition and Analysis}
The sample is imaged on an inverted microscope using Differential Interference
Contrast (DIC) optics and a 1.3 NA 100x oil-objective (Olympus).  The tethered particle's
motion was captured using an Andor Ixon camera.  Each pixel dimension corresponds to 102\thinspace{}nm in the sample plane.  Image transfer and storage was either controlled through Ixon
software (Andor Technology) or custom Matlab code (all of our Matlab acquisition and analysis code is available upon request).  The former recorded 8-14 bits per pixel, while the latter captured 14 bits per pixel.  However, a comparison of the capture methods showed insignificant differences (data not shown).  Care was taken to ensure that the image intensity exhibited broad dynamic range without saturation.  Some data was obtained using a Matlab-based autofocus routine that interfaced with a Prior controller.  However, for acquisition times shorter than five minutes, the paraxial drift was small and autofocus was not needed.

The first step in analyzing TPM data is to compute trajectories for every tethered particle.  The particle's X and Y displacement as a function of time was extracted from the raw data using a cross-correlation tracking algorithm \citep{Gelles1988}.  Such raw positional data are subject to  a slow drift due to vibrations of the experimental apparatus.  A drift correction is then applied using high pass first-order Butterworth filter at cutoff frequency 0.05\thinspace{}Hz~\citep{Vanzi2006}.



\section{Acknowledgements}
The senior authors gratefully acknowledge Nick Cozarelli's direct and indirect influence on our work.  We especially recall Nick's tactful, wise counsel at a time when one of us was an embryonic biological physicist with an
interesting, but poorly presented, idea. Multiplied manyfold, such attentions have shaped a generation of researchers.

We are grateful to  generous colleagues who have advised us on many aspects
of this work, including: Meredith
Betterton, David Dunlap, Laura Finzi, Arivalagan Gajraj, Jeff Gelles, Jan\'e Kondev, Chris Meiners, Keir
Neuman, Matthew Pennington, Tom Perkins, Bob Schleif, Kevin Towles.

\bibliography{calibrationBib}



\end{document}